# Construction of potential functions associated with a given energy spectrum

A. D. Alhaidari

*Saudi Center for Theoretical Physics, P.O. Box 32741, Jeddah 21438, Saudi Arabia*

**Abstract:** Using a formulation of quantum mechanics based on orthogonal polynomials in the energy and physical parameters, we present a method that gives the class of potential functions for exactly solvable problems corresponding to a given energy spectrum. In this work, we study the class of problems with a mix of continuous and discrete energy spectrum that are associated with the continuous dual Hahn polynomial. These include the one-dimensional logarithmic potential and the three-dimensional Coulomb plus linear potential.

***This work is dedicated to the memory of my friend, colleague and collaborator, the late Mohammed S. Abdelmonem.***

## 1. Introduction

The "Tridiagonal Representation Approach" (TRA) is an algebraic method for solving the quantum mechanical problem [1]. It is based on the theory of orthogonal polynomials and results in a class of exactly solvable problems that is larger than the conventional class. Inspired by the TRA and based on the connection between the asymptotics of orthogonal polynomials and scattering [2-4], we developed a formulation of quantum mechanics based not on potential functions but rather on orthogonal polynomials in the energy and physical parameters [5-8]. All structural and dynamical information about the physical system (e.g., bound states energies, scattering phase shift, density of states, etc.) are obtained analytically from the properties of these orthogonal polynomials (e.g., their weight function, zeros, asymptotics, etc.). In general, the energy spectrum of a quantum mechanical system is a mix of continuous scattering states and discrete bound states and we can write the total space-time wavefunction (in the atomic units $\hbar = m = 1$) as

$$\Psi(x,t) = \int e^{-iEt}\psi_E(x)dE + \sum_k e^{-iE_k t}\psi_k(x), \tag{1}$$

where $\{E_k\}$ is a finite or countably infinite discrete set of bound state energies. The Schrödinger wave equation, $i\frac{\partial\Psi}{\partial t} = H\Psi$, gives $H\psi_E = E\psi_E$ and $H\psi_k = E_k\psi_k$, where $H$ is the Hamiltonian operator. In this formulation, the wavefunction components $\psi_E(x)$ and $\psi_k(x)$ are taken as elements in an infinite dimensional vector space spanned by a complete set of square integrable basis functions $\{\phi_n(x)\}$ and we write these components as follows [5-8]

$$\psi_E(x) = \sqrt{\sigma(E)}\sum_n P_n(s)\,\phi_n(x), \tag{2a}$$

$$\psi_k(x)=\sqrt{\omega(s_k)}\sum_n P_n(s_k)\,\phi_n(x)\,, \tag{2b}$$

where $\{P_n(s)\}$ are orthogonal polynomials with a mixed spectrum. The polynomial argument s is related to the energy by the measure relation $\sigma(E)dE=\rho(s)ds$ where $\rho(s)$ is the continuous component of the positive definite weight function for $\{P_n(s)\}$ whereas $\omega(s_k)$ is the discrete component of the weight function. These polynomials (in their orthonormal version) satisfy the following generalized orthogonality relation

$$\int \rho(s)P_n(s)P_m(s)\,ds+\sum\nolimits_k \omega(s_k)P_n(s_k)P_m(s_k)=\delta_{n,m}\,. \tag{3}$$

In fact, inserting (2a) and (2b) in (1) and using this orthogonality along with the completeness of the basis, one can show that the wavefunction and its conjugate are space-correlated over all times as follows

$$\int \Psi(x,t)\overline{\Psi}(y,t)dt=\delta(x-y)\,, \tag{4}$$

where the conjugate wavefunction, $\overline{\Psi}(x,t)$, is obtained from (1) and (2) by the replacements $\mathrm{i}\mapsto-\mathrm{i}$ and $\phi_n(x)\mapsto\overline{\phi}_n(x)$ with $\langle\phi_n|\overline{\phi}_m\rangle=\langle\overline{\phi}_n|\phi_m\rangle=\delta_{n,m}$ and $\sum_{n=0}^{\infty}\phi_n(x)\overline{\phi}_n(y)=\delta(x-y)$.

Additionally, the polynomials $\{P_n(s)\}$ satisfy a three-term recursion relation whose symmetric version reads as follows

$$s\,P_n(s)=a_nP_n(s)+b_{n-1}P_{n-1}(s)+b_nP_{n+1}(s)\,, \tag{5}$$

where $n=1,2,3,...$ The coefficients $\{a_n,b_n\}$ depend on n and the physical parameters but are independent of the polynomial argument s and such that $b_n^2>0$ for all n. This recursion relation gives $P_n(s)$ as a polynomial of degree n in s and determines all of them explicitly starting with the initial two, $P_0(s)$ and $P_1(s)$. Normally, one takes $P_0(s)=1$, $P_1(s)=(s-a_0)/b_0$ and they are referred to as "polynomials of the first kind". An example of such polynomial with mixed spectrum is the Wilson polynomial, which belongs to the Askey scheme of hypergeometric orthogonal polynomials [9]. The asymptotics ($n\to\infty$) of $P_n(s)$ is a sinusoidal function in n and it gives the spectrum formula for $P_n(s)$ as an infinite (or finite) discrete set $\{s_k\}$ at which the amplitude of the sinusoidal oscillation vanishes. The scattering phase shift is obtained from the argument of the sinusoidal function [5-8].

Therefore, given the set of polynomials $\{P_n(s)\}$ one only needs to choose the basis elements $\{\phi_n(x)\}$ for a full determination of the wavefunction (1). Once this is done, the system becomes completely defined as it is well demonstrated in several studies (see, for examples, the work in [5-8]). Note that all of this physics is acquired despite the fact that no reference has ever been made to any potential function. That is, we did not require that the Hamiltonian be represented as the sum of a kinetic energy operator and a potential function. This, in fact, opens the door to identifying a larger class of exactly (analytically) integrable quantum systems without worrying whether a potential function could be realized analytically or not. However, researchers who are accustomed to working in the standard formulation of quantum mechanics may still desire to identify the corresponding potential function. Establishing this correspondence in illustrative examples is precisely the objective of this work. If successful, then it could be considered as one of the solutions to the inverse problem. That is, constructing the potential function using knowledge of the energy spectrum data. As we shall find out, however, this solution is not unique due to an equivalence generated by similarity transformations of the corresponding Hamiltonian matrix. We should also note that insisting that the Hamiltonian be represented as the sum of a kinetic energy operator and a potential function may lead to potential functions that could only be realized numerically. In fact, all potential functions associated with the *new* exactly integrable quantum systems presented in this work can only be obtained numerically for a given set of physical parameters. Nonetheless, we do succeed in making attempts to have excellent fit of the resulting potentials to some analytic functions. Based on these empirical results, we conjecture that the following potential functions have exact solution as presented in subsections 3.1 and 3.3, respectively:

(1) In 3D with spherical symmetry: $V(r)=\dfrac{V_0/\lambda}{r}+\lambda V_1 r$. (6a)

(2) On the positive real line in 1D: $V(x)=V_0\ln(1+\lambda x)$. (6b)

where $\lambda$ is a positive parameter of inverse length dimension and $\{V_i\}$ are real potential parameters.

In section 2, we start with a given orthogonal polynomial $P_n(s)$ of a mixed spectrum that results in an energy spectrum formula whose structure is common to a large class of exactly solvable systems. Subsequently, we develop a procedure to derive the corresponding potential functions for other integrable systems each of which is associated with a given basis sets $\{\phi_n(x)\}$. In section 3, we give four illustrative examples using variants of a special basis set called the "Laguerre basis". We end the work in section 4 with a conclusion, discussion of our findings and an outlook for further studies.

## 2. Formulation

Substituting the wavefunction expansion (2) in the time-independent wave equation $H|\psi_E\rangle=E|\psi_E\rangle$ and projecting from left by $\langle\phi_n|$, we obtain the following equivalent eigenvalue matrix wave equation

$$\sum_m H_{n,m}|P_m\rangle=E\sum_m\Omega_{n,m}|P_m\rangle, \tag{7}$$

where $H_{n,m}=\langle\phi_n|H|\phi_m\rangle$ are elements of the matrix representation the Hamiltonian in the basis $\{\phi_n\}$ and $\Omega$ is the overlap matrix among the basis elements, $\Omega_{n,m}=\langle\phi_n|\phi_m\rangle$. For simplicity, we consider only orthonormal basis set where $\Omega_{n,m}=\delta_{n,m}$. We rewrite the recursion relation (5) as the matrix equation $s|P\rangle=\Sigma|P\rangle$, where $\Sigma$ is the tridiagonal symmetric matrix with elements

$$\Sigma_{n,m}=a_n\delta_{n,m}+b_{n-1}\delta_{n,m+1}+b_n\delta_{n,m-1}. \tag{8}$$

If we write the energy spectrum formula as $E=F(s)$, where $F(s)$ is an entire function, then we can formally write $E|P\rangle=F(s)|P\rangle=F(\Sigma)|P\rangle$. Comparing this to Eq. (7) with $\Omega=1$, which

reads $H|P\rangle = E|P\rangle$, we conclude that $E$ is a common eigenvalue for the two matrices $H$ and $F(\Sigma)$. Consequently, these two matrices are related by a similarity transformation and we can write

$$H = \Lambda\left[F(\Sigma)\right]\Lambda^{-1}, \tag{9}$$

where $\Lambda$ is the non-singular similarity transformation matrix. If $F(\Sigma)$ is a symmetric tridiagonal matrix, then we choose $\Lambda$ as the identity matrix, $\Lambda = 1$. However, if it is not, then we choose $\Lambda$ such that $\Lambda\left[F(\Sigma)\right]\Lambda^{-1}$ becomes tridiagonal and symmetric. One such matrix is the Householder transformation matrix (see section 11.3 of Ref. [10]), which has the property $\Lambda^{\mathrm{T}} = \Lambda^{-1}$.

In this work, we choose the following energy spectrum, which is common to a large class of exactly solvable problems in the standard formulation of quantum mechanics

$$E_k = -\frac{\lambda^2}{2}\left(k+\mu\right)^2, \tag{10}$$

where $\lambda$ is a positive real parameter of inverse length dimension and $\mu$ is a negative dimensionless real parameter. This energy spectrum is finite with $k = 0,1,..,N$ and $N$ is the largest integer less than or equal to $-\mu$. Table 1 is a list of exactly solvable problems in the conventional formulation of quantum mechanics with a mix of continuous and discrete energy spectra such that the latter is finite and has the form (10) (see, for example, Ref. [11]). However, we will show below that there are many other integrable systems with such an energy spectrum whose wavefunctions are written analytically using the explicit mathematical formulation given by Eq. (1) and Eq. (2). To make this point (numerous systems with the same energy spectrum formula) abundantly clear, we make the following argument where we draw from analogy to vector calculus:

In physics, we are accustomed to writing vector quantities (e.g., force, velocity, electric field, etc.) in terms of their components in some conveniently chosen vector space. For example, the force $\vec{F}$ in three dimensional space is written in Cartesian coordinates as $\vec{F} = f_x\hat{x} + f_y\hat{y} + f_z\hat{z}$, where $\left\{f_x, f_y, f_z\right\}$ are the components/projections of the force along the unit vectors $\left\{\hat{x},\hat{y},\hat{z}\right\}$. These components contain all physical information about the quantity whereas the unit vectors (basis) are dummy, but must form a complete set to allow for a faithful physical representation. For example, to fully represent the force in 3D space, it is not sufficient to write it in terms of only two of its components as $\vec{F} = f_x\hat{x} + f_y\hat{y}$. Now, we can as well write the same force in another coordinates, say the spherical coordinates with basis $\left\{\hat{r},\hat{\theta},\hat{\varphi}\right\}$, as $\vec{F} = f_r\hat{r} + f_\theta\hat{\theta} + f_\varphi\hat{\varphi}$, where $\left\{f_r, f_\theta, f_\varphi\right\}$ are the new components/projections that differ from $\left\{f_x, f_y, f_z\right\}$ but contain the same physical information. On the other hand, keeping the same components while changing the basis by writing $\vec{F}' = f_x\hat{r} + f_y\hat{\theta} + f_z\hat{\varphi}$ will certainly produce a physically different vector quantity. Therefore, the number of distinct physical systems with same projections is equal to the number of proposed bases. This analysis can be extended to infinite dimensional vector space by writing $\vec{F} = \sum_n f_n\hat{x}_n$ where the basis (unit vectors) $\left\{\hat{x}_n\right\}$ is chosen conveniently depending on the symmetry of the problem (e.g., rectangular, spherical, cylindrical, elliptical, etc.). In this analogy, we make the map $\vec{F} \mapsto \psi$, $\left\{\hat{x}_n\right\} \mapsto \left\{\phi_n\right\}$, $\left\{f_n\right\} \mapsto \sqrt{\rho}\left\{P_n\right\}$ and the relevant point in the argument is that changing $\{\phi_n(x)\}$ while keeping the same $\{P_n(s)\}$ leads to a physically different quantum systems while maintaining exact (analytical) knowledge of both.

Now, one of the hypergeometric orthogonal polynomials in the Askey scheme (but not the only one) that has a mix of continuous and finite discrete spectrum with a spectrum formula similar to (10) is the three-parameter continuous dual Hahn polynomial $S_n^\mu(z^2;a,b)$ whose orthonormal version could be written as follows

$$S_n^\mu(z^2;a,b)=\sqrt{\frac{(\mu+a)_n(\mu+b)_n}{n!(a+b)_n}}\;{}_3F_2\left(\left.\begin{matrix}-n,\mu+\mathrm{i}z,\mu-\mathrm{i}z\\ \mu+a,\mu+b\end{matrix}\right|1\right), \tag{11}$$

where ${}_3F_2\left(\left.\begin{matrix}a,b,c\\ d,e\end{matrix}\right|x\right)=\sum_{n=0}^{\infty}\frac{(a)_n(b)_n(c)_n}{(d)_n(e)_n}\frac{x^n}{n!}$ is the generalized hypergeometric function and $(a)_n=a(a+1)(a+2)...(a+n-1)=\frac{\Gamma(n+a)}{\Gamma(a)}$. The properties of this polynomial are given in Appendix B of Ref. [8]. Most importantly is that if $\mathrm{Re}(\mu,a,b)>0$ with non-real parameters occurring in conjugate pairs then $S_n^\mu(z^2;a,b)$ is a polynomial of degree $n$ in $z^2$ with a purely continuous spectrum on the positive real line, $z\geq 0$. In this case, the orthogonality of these polynomials reads as follows (see Appendix B in [8])

$$\int_0^\infty \rho(z)S_n^\mu(z^2;a,b)S_m^\mu(z^2;a,b)dz=\delta_{nm},\quad \text{where} \tag{12a}$$

$$\rho(z)=\frac{1}{2\pi}\frac{\left|\Gamma(\mu+\mathrm{i}z)\Gamma(a+\mathrm{i}z)\Gamma(b+\mathrm{i}z)/\Gamma(2\mathrm{i}z)\right|^2}{\Gamma(\mu+a)\Gamma(\mu+b)\Gamma(a+b)}. \tag{12b}$$

On the other hand, if the parameters are such that $\mu<0$ and $a+\mu$, $b+\mu$ are positive or a pair of complex conjugates with positive real parts, then the polynomial will have a continuous spectrum on $z\geq 0$ as well as a finite size discrete spectrum with $z^2<0$. In this case, the polynomial satisfies a generalized orthogonality relation similar to (3) and given by Eq. (B3) in [8] where the discrete weight function reads

$$\omega(z_k)=-2\frac{\Gamma(a-\mu)\Gamma(b-\mu)}{\Gamma(a+b)\Gamma(1-2\mu)}\left[\left(k+\mu\right)\frac{(\mu+a)_k(\mu+b)_k(1-k-2\mu)_k}{(a-\mu-k)_k(b-\mu-k)_k\,k!}\right], \tag{13}$$

where we have used $(-a)_n=(-1)^n(a-n+1)_n$. The symmetric three-term recursion relation satisfied by $S_n^\mu(z^2;a,b)$ is given by Eq. (B4) in Appendix B of [8]. Comparing that to Eq. (5) above makes $s=z^2$ and gives the following elements of the tridiagonal symmetric matrix $\Sigma$

$$\begin{aligned}\Sigma_{n,m}=&\left[(n+\mu+a)(n+\mu+b)+n(n+a+b-1)-\mu^2\right]\delta_{n,m}\\ &-\sqrt{n(n+a+b-1)(n+\mu+a-1)(n+\mu+b-1)}\,\delta_{n,m+1}\\ &-\sqrt{(n+1)(n+a+b)(n+\mu+a)(n+\mu+b)}\,\delta_{n,m-1}\end{aligned} \tag{14}$$

Moreover, the spectrum formula of $S_n^\mu(z^2;a,b)$, which is given by Eq. (B8) in [8], reads

$$z_k^2=-(k+\mu)^2. \tag{15}$$

Comparing this with Eq. (10) gives the energy spectrum formula $E = F(z^2) = \frac{1}{2}\lambda^2 z^2$. Therefore, the matrix representation of the Hamiltonian operator (9) becomes $H = \frac{1}{2}\lambda^2\,\Sigma$. Now, since $F(\Sigma)$ is tridiagonal, then our choice for the similarity transformation in (9) is $\Lambda = 1$ .

In the standard formulation of quantum mechanics, the Hamiltonian *H* is the sum of the kinetic energy operator *T* and the potential function *V*. Thus, $V = H - T$ and to obtain the matrix representation of the potential in the basis $\{\phi_n\}$ with *H* given by Eq. (9), we need to compute the matrix elements of *T* in this basis. Now, *T* is usually a well-known differential operator in configuration space that depends only on the number of dimensions but is independent of the type of interaction potential. For example, in one dimension with coordinate *x* and adopting the atomic units $\hbar = m = 1$, $T = -\frac{1}{2}\frac{d^2}{dx^2}$. In three dimensions with spherical symmetry and radial coordinate *r*, $T = -\frac{1}{2}\frac{d^2}{dr^2} + \frac{\ell(\ell+1)}{2r^2}$ where $\ell$ is the angular momentum quantum number. Therefore, the action of *T* on the given basis elements $\{\phi_n\}$ could be derived and so too its matrix elements. Consequently, with the matrices *H* and *T* being determined, we can obtain the potential matrix *V* in the basis $\{\phi_n\}$ as $H - T$. Now, with the matrix representation of the potential being identified and the basis set in which it is calculated is given, we can then use one of four methods developed in section 3 of ref. [12] to construct the potential function. Our choice is the second method, which, for ease of reference, is presented here in Appendix A. We illustrate the procedure by giving a set of examples using a special basis called the "Laguerre basis" whose elements are

$$\phi_n(x) = A_n y^{\alpha} e^{-\beta y} L_n^{\nu}(y), \tag{16}$$

where $L_n^{\nu}(y)$ is the Laguerre polynomial whose dimensionless argument is a function of the configuration space coordinate such that $y(x) \geq 0$ and $A_n$ is a proper normalization constant. The dimensionless real parameters $\{\alpha, \beta, \nu\}$ are such that $\beta > 0$ and $\nu > -1$. Table 2 gives the parameters of orthonormal set of Laguerre bases that are relevant to the examples in section 3. To verify orthonormality of this set, one uses the orthogonality of the Laguerre polynomials

$$\int_0^{\infty} y^{\nu} e^{-y} L_n^{\nu}(y) L_m^{\nu}(y) dy = \frac{\Gamma(n+\nu+1)}{\Gamma(n+1)}\delta_{n,m}, \tag{17}$$

in addition to the integral measure $\int_{x_-}^{x_+} dx = \int_0^{\infty} \frac{dy}{y'}$, where $x_{\pm}$ are the boundaries of configuration space and $y' = dy/dx$. To demonstrate the accuracy of our scheme, we choose one of the examples to correspond to an exactly solvable problem from the conventional class of potentials shown in Table 1 and compare with our results.

## 3. Illustrative examples

In the following four subsections, we present examples of systems with a mixed energy spectrum where the number of discrete bound states is finite and the total wavefunction is given by Eq. (1) and Eq. (2). The energy polynomial in these expressions is $P_n(z) = S_n^{\mu}(z^2; a, b)$ with $E=\frac{1}{2}\lambda^2 z^2$. The elements of the basis $\{\phi_n(x)\}$ are given by Eq. (16) with the parameters shown in the entries of Table 2. The matrix representation of the kinetic energy is obtained by calculating the action of the kinetic energy differential operator on the elements of the basis as $T|\phi_m\rangle$ then computing the matrix elements by projecting from left with $\langle\phi_n|$ then evaluating the integral $T_{n,m}=\langle\phi_n|T|\phi_m\rangle$. In the calculation of $T|\phi_m\rangle$, one needs the following differential equation, differential property and recursion relation of the Laguerre polynomials

$$\left[y\frac{d^2}{dy^2}+\left(\nu+1-y\right)\frac{d}{dy}+n\right]L_n^\nu(y)=0\,, \tag{18a}$$

$$y\frac{d}{dy}L_n^\nu(y)=n\,L_n^\nu(y)-(n+\nu)L_{n-1}^\nu(y)\,, \tag{18b}$$

$$y\,L_n^\nu(y)=(2n+\nu+1)L_n^\nu(y)-(n+\nu)L_{n-1}^\nu(y)-(n+1)L_{n+1}^\nu(y)\,. \tag{18c}$$

Moreover, in the process of calculating the matrix elements $\langle\phi_n|T|\phi_m\rangle$, one encounters integrals of the form

$$A_nA_m\int_0^\infty y^{\nu\pm k}e^{-(1+\tau)y}L_n^\nu(y)L_m^\nu(y)dy:=F_{n,m}^{\pm}(k,\tau)\,, \tag{19}$$

where *k* is a non-negative integer, $\tau\geq 0$ and $A_n=\sqrt{\frac{\Gamma(n+1)}{\Gamma(n+\nu+1)}}$. Such integrals are evaluated using results from the work of Srivastava *et al.* [13] as outlined in Appendix B. Finally, we write the potential matrix as $V=H-T$. With this matrix and the basis (16), the technique described in Appendix A gives the potential function for a selected set of physical parameters.

**3.1 *The case*** $y(x)=\lambda x$ and $(2\alpha,2\beta)=(\nu,1)$:

This case corresponds to the basis with parameters shown in the first row of Table 2. Using the properties of the Laguerre polynomial given by Eq. (18a)-(18c), we obtain the following action of the kinetic energy operator in 3D with spherical symmetry on the basis

$$\begin{aligned}-\frac{2}{\lambda^2}T|\phi_n(x)\rangle&=\left[\frac{d^2}{dy^2}-\frac{\ell(\ell+1)}{y^2}\right]|\phi_n(x)\rangle=A_ny^{\frac{1}{2}\nu}e^{-\frac{1}{2}y}\left[\left(\frac{d}{dy}+\frac{\nu/2}{y}-\frac{1}{2}\right)^2-\frac{\ell(\ell+1)}{y^2}\right]|L_n^\nu(y)\rangle\\&=A_ny^{\frac{1}{2}\nu}e^{-\frac{1}{2}y}\left[\frac{d^2}{dy^2}+\left(\frac{\nu}{y}-1\right)\frac{d}{dy}-\frac{\nu/2}{y}+\frac{\frac{1}{4}(\nu-1)^2-\left(\ell+\frac{1}{2}\right)^2}{y^2}+\frac{1}{4}\right]|L_n^\nu(y)\rangle\\&=A_ny^{\ell+1}e^{-\frac{1}{2}y}\left\{\left[-\frac{n}{y^2}-\frac{n+\ell+1}{y}+\frac{1}{4}\right]|L_n^{2\ell+2}(y)\rangle+\frac{n+2(\ell+1)}{y^2}|L_{n-1}^{2\ell+2}(y)\rangle\right\}\end{aligned} \tag{20}$$

where in the last step we have chosen the basis parameter $\nu=2(\ell+1)$ to eliminate the most singular $y^{-2}$ term in the previous step. Using (20) to evaluate the matrix elements $\langle\phi_n|T|\phi_m\rangle$, we encounter integrals in the form given by Eq. (19). One of them is $F_{n,m}^{+}(0,0)$, which is equal to $\delta_{n,m}$. However, we also encounter the integrals $F_{n,m}^{-}(1,0)$ and $F_{n,m}^{-}(2,0)$, which are evaluated in Appendix B as given by Eq. (B4). After some manipulations and rearrangement, we obtain the following matrix representation of the kinetic energy operator in the basis (16)

$$T_{n,m}=\frac{\lambda^2}{4}\begin{cases}\sqrt{(n+1)_{2\ell+2}/(m+1)_{2\ell+2}}\left(1+\frac{2n}{2\ell+3}\right) & ,m>n\\ \sqrt{(m+1)_{2\ell+2}/(n+1)_{2\ell+2}}\left(1+\frac{2m}{2\ell+3}\right) & ,m<n\\ \frac{1}{2}+\frac{2n}{2\ell+3} & ,m=n\end{cases} \tag{21}$$

Using this matrix and $H=\frac{1}{2}\lambda^2\Sigma$, we obtain the potential matrix as $H-T$. Then, we employ the method described in Appendix A that uses the potential matrix and the basis (16) to produce the potential function shown as solid line in Fig. 1 (after adding the orbital term $\frac{\ell(\ell+1)}{2x^2}$) for a given set of physical parameters $\{\lambda,\mu,\ell\}$. We also superimpose the energy spectrum levels given by Eq. (10) on the same plot. By empirical trials, we were able to obtain a perfect fit to the effective potential $V_{eff}(x)=\frac{\ell(\ell+1)}{2x^2}+\lambda^{-1}V_0\,x^{-1}+\lambda V_1\,x+V_2$ for a given set of parameters $V_0$, $V_1$ and $V_2$. This is shown as the dotted line in Fig. 1.

**3.2 *The case*** $y(x)=(\lambda x)^2$ **and** $\left(2\alpha,2\beta\right)=\left(\nu+\frac{1}{2},1\right)$:

This case corresponds to the basis with parameters shown in the second row of Table 2. Using the properties of the Laguerre polynomial given by Eq. (18a)-(18c), we obtain the following action of the kinetic energy operator in 3D with spherical symmetry on the basis elements (16)

$$\begin{aligned}-\frac{1}{2\lambda^2}T\left|\phi_n(x)\right\rangle &=\left[y\frac{d^2}{dy^2}+\frac{1}{2}\frac{d}{dy}-\frac{\ell(\ell+1)}{4y}\right]\left|\phi_n(x)\right\rangle\\ &=\sqrt{2}A_n y^{\frac{1}{2}(\nu+\frac{1}{2})}e^{-\frac{1}{2}y}\left[y\frac{d^2}{dy^2}+\left(\nu+1-y\right)\frac{d}{dy}+\frac{\nu^2-\left(\ell+\frac{1}{2}\right)^2}{4y}+\frac{y}{4}-\frac{\nu+1}{2}\right]\left|L_n^{\nu}(y)\right\rangle\\ &=\frac{1}{\sqrt{2}}A_n y^{\frac{1}{2}(\ell+1)}e^{-\frac{1}{2}y}\left[\frac{y}{2}-\left(2n+\ell+\frac{3}{2}\right)\right]\left|L_n^{\ell+\frac{1}{2}}(y)\right\rangle\end{aligned} \tag{22}$$

where we have chosen the basis parameter $\nu=\ell+\frac{1}{2}$ to eliminate the most singular $y^{-1}$ term in the middle step. Using the recursion relation and orthogonality of the Laguerre polynomials given by Eq. (18c) and Eq. (17), we obtain a tridiagonal symmetric matrix representation for the kinetic energy operator whose elements are as follows

$$T_{n,m}=\frac{\lambda^2}{2}\left[\left(2n+\ell+\frac{3}{2}\right)\delta_{n,m}+\sqrt{n\left(n+\ell+\frac{1}{2}\right)}\,\delta_{n,m+1}+\sqrt{(n+1)\left(n+\ell+\frac{3}{2}\right)}\,\delta_{n,m-1}\right]. \tag{23}$$

Using this matrix and $H=\frac{1}{2}\lambda^2\Sigma$, we obtain the potential matrix as $V=H-T$ and employ the method outlined in Appendix A to produce the potential function shown as solid line in Fig. 2 (after adding the orbital term $\frac{\ell(\ell+1)}{2x^2}$) for a given set of physical parameters $\{\lambda,\mu,\ell\}$. We also superimpose the energy spectrum levels given by Eq. (10) on the same plot. By empirical trials, we were able to make a perfect fit to the effective potential $V_{eff}(x)=\frac{\ell(\ell+1)}{2x^2}+\lambda^2V_0x^2+V_1$ for a given set of potential parameters $V_0$ and $V_1$. This is shown as the dotted line in Fig. 2.

**3.3 *The case*** $y(x)=\gamma\ln\left(1+\lambda x\right)$ and $\left(2\alpha,2\beta\right)=\left(\nu,1+\gamma^{-1}\right)$:

This case corresponds to the basis with the parameters shown in the third row of Table 2. Employing the properties of the Laguerre polynomial given by Eq. (18a)-(18c), we obtain the following action of the kinetic energy operator in 1D, $T=-\frac{1}{2}\frac{d^2}{dx^2}$, on the basis

$$\begin{aligned}
-\frac{2}{(\gamma\lambda)^2}T\left|\phi_n(x)\right\rangle &= e^{-2\gamma^{-1}y}\left(\frac{d^2}{dy^2}-\frac{1}{\gamma}\frac{d}{dy}\right)\left|\phi_n(x)\right\rangle=\sqrt{\gamma}A_n y^{\frac{1}{2}\nu}e^{-\frac{1}{2}(1+5\gamma^{-1})y}\\
&\times\left[-\left(\frac{1}{y}+\frac{2}{\gamma}\right)\frac{d}{dy}+\frac{\nu(\nu-2)}{4y^2}-\frac{2n+\nu(1+2\gamma^{-1})}{2y}+\frac{1}{4}\left(1+\gamma^{-1}\right)\left(1+3\gamma^{-1}\right)\right]\left|L_n^\nu(y)\right\rangle\\
&=\sqrt{\gamma}A_n y^{\frac{1}{2}\nu}e^{-\frac{1}{2}(1+5\gamma^{-1})y}\left\{\left[-\frac{2n+\nu-\nu^2/2}{2y^2}-\left(1+2\gamma^{-1}\right)\frac{2n+\nu}{2y}+\frac{1}{4}\left(1+\gamma^{-1}\right)\left(1+3\gamma^{-1}\right)\right]\left|L_n^\nu(y)\right\rangle\right.\\
&\qquad\left.+\left(n+\nu\right)\left(\frac{1}{y^2}+\frac{2\gamma^{-1}}{y}\right)\left|L_{n-1}^\nu(y)\right\rangle\right\}
\end{aligned}\tag{24}$$

Using this to evaluate the matrix elements $\left\langle\phi_n\right|T\left|\phi_m\right\rangle$, we encounter integrals similar to those given by Eq. (19). Evaluating these as shown in Appendix B, we obtain the following matrix representation of the kinetic energy in the basis (16)

$$\begin{aligned}
\frac{4}{(\gamma\lambda)^2}T_{n,m} &= -\frac{1}{4}\left(1+\tfrac{1}{2}\tau\right)\left(1+\tfrac{3}{2}\tau\right)F_{n,m}^-(0,\tau)+\left(1+\tau\right)\left(m+\tfrac{\nu}{2}\right)F_{n,m}^-(1,\tau)\\
&+\tfrac{1}{2}\left(2m+\nu-\tfrac{1}{2}\nu^2\right)F_{n,m}^-(2,\tau)-\left(m+\nu\right)\left[\tau F_{n,m-1}^-(1,\tau)+F_{n,m-1}^-(2,\tau)\right]+n\leftrightarrow m
\end{aligned}\tag{25}$$

where $\tau=2\gamma^{-1}$ †. The matrices $T$ and $H=\frac{1}{2}\lambda^2\Sigma$ give the potential matrix, which is then utilized by the method outlined in Appendix A to produce the potential function shown as solid line in Fig. 3 for a given set of physical parameters $\left\{\lambda,\mu,\gamma\right\}$. By empirical trials, we obtained a perfect fit to the potential function $V(x)=V_0\ln(1+\lambda x)+V_1$ for a given set of parameters $V_0$ and $V_1$. This is shown as the dotted line in Fig. 3.

**3.4 *The case*** $y(x)=e^{\lambda x}$ and $\left(2\alpha,2\beta\right)=\left(\nu+1,1\right)$:

This case corresponds to the basis with parameters shown in the fourth row of Table 2. Moreover, this choice of basis and polynomial $S_n^\mu(z^2;a,b)$ corresponds to the 1D Morse problem shown in Table 1 with $\mu=A+\frac{1}{2}$ (see, for example, section 4.1 in Ref. [8]). We will show that we our scheme reproduces these exact results. Utilizing the properties of the Laguerre polynomial, we obtain the following action of the kinetic energy operator in 1D, $T=-\frac{1}{2}\frac{d^2}{dx^2}$, on the basis

---

† We can simplify Eq. (25) by eliminating the singular third term if we choose $\nu=1+\sqrt{1+4m}$, which makes the index of the Laguerre polynomial depend on its degree. However, this leads to an expression for $F_{n,m}^-(k,\tau)$ that is more elaborate than (B3).

$$-\frac{2}{\lambda^2}T\left|\phi_n(x)\right\rangle = \left(y^2\frac{d^2}{dy^2}+y\frac{d}{dy}\right)\left|\phi_n(x)\right\rangle$$

$$= A_n y^{\frac{1}{2}(\nu+1)}e^{-\frac{1}{2}y} y\left[y\frac{d^2}{dy^2}+\left(\nu+2-y\right)\frac{d}{dy}+\frac{\left(\nu+1\right)^2}{4y}+\frac{y}{4}-\frac{\nu+2}{2}\right]\left|L_n^\nu(y)\right\rangle \tag{26}$$

$$= A_n y^{\frac{1}{2}(\nu+1)}e^{-\frac{1}{2}y}\left\{\left[\frac{y^2}{4}-\left(2n+\nu+2\right)\frac{y}{2}+n+\frac{(\nu+1)^2}{4}\right]\left|L_n^\nu(y)\right\rangle-\left(n+\nu\right)\left|L_{n-1}^\nu(y)\right\rangle\right\}$$

Utilizing the recursion relation and orthogonality of the Laguerre polynomials, we obtain a penta-diagonal symmetric matrix representation for the kinetic energy operator whose elements are as follows

$$\frac{4}{\lambda^2}T_{n,m} = -\frac{1}{2}\left(J^2\right)_{n,m}+\left[\left(2n+\nu+1\right)^2+\tfrac{1}{2}\left(1-\nu^2\right)\right]\delta_{n,m}$$
$$-\left(2n+\nu\right)\sqrt{n\left(n+\nu\right)}\,\delta_{n,m+1}-\left(2n+\nu+2\right)\sqrt{\left(n+1\right)\left(n+\nu+1\right)}\,\delta_{n,m-1} \tag{27}$$

where $J$ is the tridiagonal symmetric matrix with elements given in Appendix B by Eq. (B2). Using the matrices $T$ and $H=\frac{1}{2}\lambda^2\,\Sigma$, we obtain the potential matrix and then employ the method in Appendix A to produce the potential function shown as solid line in Fig. 4 for a given set of values of the physical parameters $\{\lambda,\mu\}$. We also superimpose on the same plot the energy levels given by Eq. (10). The figure shows an excellent agreement with the Morse potential $V(x)=\frac{1}{2}\lambda^2\left(\frac{1}{4}e^{2\lambda x}+Ae^{\lambda x}\right)$, which is shown as the dotted curve in Fig. 4.

## 4. Conclusion and discussion

In this work, we have shown that the formulation of quantum mechanics based not on potential functions but orthogonal polynomials in the energy and physical parameters lead to a larger class of exactly solvable problems. This was demonstrated for a given energy spectrum formula that is common to various exactly solvable potentials in the conventional class. The process starts by identified polynomials $\{P_n(s)\}$ with the same structure of spectrum formula. Using those together with a properly chosen basis set $\{\phi_n(x)\}$, we write the total wavefunction as given by Eq. (1) and Eq. (2). The kinetic energy matrix representation is calculated by computing the action of the kinetic energy operator on the basis. The Hamiltonian matrix is obtained using the symmetric three-term recursion relation and the spectrum formula of the selected polynomials $\{P_n(s)\}$. Subsequently, the potential function is derived by employing several methods (one of them is outlined in Appendix A) that utilize the matrix elements of the potential $V=H-T$ and the basis set. We conjectured that the linear potential (6a) and the logarithmic potential (6b) have exact solutions whose energy spectra are given by Eq. (10) and wavefunctions are written as shown by Eq. (2) with $P_n(z)=S_n^\mu(z^2;a,b)$ and $\phi_n(x)$ as given by Eq. (16). The success of this technique constitutes a solution to the inverse problem where the energy spectrum data is used to construct the interaction potential function.

We plan to follow this work by another addressing the same energy spectrum formula (10) but for a different energy polynomial and another set of basis. The polynomial will be the Wilson polynomial whose orthonormal version reads as follows

$$W_n^{\mu}(z^2;\nu;a,b) == \sqrt{\left(\frac{2n+\mu+\nu+a+b-1}{n+\mu+\nu+a+b-1}\right)\frac{(\mu+a)_n(\mu+b)_n(\mu+\nu)_n(\mu+\nu+a+b)_n}{(\nu+a)_n(\nu+b)_n(a+b)_n n!}} \\ \times {}_4F_3\left(\begin{matrix}-n,n+\mu+\nu+a+b-1,\mu+iz,\mu-iz\\ \mu+\nu,\mu+a,\mu+b\end{matrix}\middle|1\right) \tag{28}$$

The properties of this polynomial could be found in section 2 and Appendix C of Ref. [8]. Moreover, the corresponding basis set, called the "Jacobi basis", has the following elements

$$\phi_n(x) = A_n(1-y)^{\alpha}(1+y)^{\beta}P_n^{(\mu,\nu)}(y), \tag{29}$$

where $P_n^{(\mu,\nu)}(y)$ is the Jacobi polynomial such that $-1 \le y(x) \le +1$ and $(\mu,\nu) > -1$. $A_n$ is a proper normalization constant. Table 3 gives the basis parameters for an orthonormal set that is relevant to the problems to be addressed.

## Appendix A: Evaluating the potential function

Let $\{V_{n,m}\}_{n,m=0}^{N-1}$ be the $N\times N$ matrix elements of the potential function $V(x)$ in a given basis set $\{\phi_n(x)\}$ and let $\{\bar{\phi}_n(x)\}$ be the conjugate basis set; that is $\langle\bar{\phi}_n|\phi_m\rangle = \langle\phi_n|\bar{\phi}_m\rangle = \delta_{nm}$. Specifically, we mean that $V_{n,m} = \langle\phi_n|V|\phi_m\rangle$. In this Appendix, we present the second out of four methods developed in section 3 of Ref. [12] to calculate the potential function using its matrix elements in the given basis set. If we write $\langle x|V|x'\rangle = V(x)\delta(x-x')$ and $\langle x|\phi_n\rangle = \phi_n(x)$, then using the completeness in configuration space, $\int|x'\rangle\langle x'|dx' = 1$, we can write

$$\langle x|V|\phi_n\rangle = \int\langle x|V|x'\rangle\langle x'|\phi_n\rangle dx' = V(x)\phi_n(x). \tag{A1}$$

On the other hand, the completeness of the basis, $\sum_n|\bar{\phi}_n\rangle\langle\phi_n| = \sum_n|\phi_n\rangle\langle\bar{\phi}_n| = I$, where $I$ is the identity, enables us to write the left side of Eq. (A1) as

$$\langle x|V|\phi_n\rangle = \sum_{m=0}^{\infty}\langle x|\bar{\phi}_m\rangle\langle\phi_m|V|\phi_n\rangle = \sum_{m=0}^{\infty}\bar{\phi}_m(x)V_{m,n} \cong \sum_{m=0}^{N-1}\bar{\phi}_m(x)V_{m,n}. \tag{A2}$$

These two equations give

$$V(x) \cong \sum_{m=0}^{N-1}\frac{\bar{\phi}_m(x)}{\phi_n(x)}V_{m,n}, \quad n = 0,1,...,N-1. \tag{A3}$$

Therefore, we need the information in only one column of the potential matrix (or one row, since $V_{n,m} = V_{m,n}$ ) and the basis set to determine $V(x)$. In particular, if we choose $n = 0$, we obtain

$$V(x) \cong \sum_{m=0}^{N-1}\frac{\bar{\phi}_m(x)}{\phi_0(x)}V_{m,0}. \tag{A4}$$

Note that our choice of basis in this work as an orthonormal set makes it self-conjugate. That is, $\overline{\phi}_n(x) = \phi_n(x)$.

## Appendix B: Evaluating the integral (19)

In this Appendix, we evaluate the integral $F^{\pm}_{n,m}(k,\tau)$ defined in Eq. (19) with *k* a non-negative integer and $\tau \geq 0$. We consider two separate cases:

(1) $F^{+}_{n,m}(k,0)$ and $\nu > -1$.

(2) $F^{-}_{n,m}(k,\tau)$ and $\nu > k-1$.

For the first case, we use the three-term recursion relation of the Laguerre polynomials, Eq. (18c), repeatedly to obtain

$$F^{+}_{n,m}(k,0) = \left(J^k\right)_{n,m}, \tag{B1}$$

where *J* is the symmetric tridiagonal matrix whose elements are

$$J_{n,m} = (2n+\nu+1)\delta_{n,m} - \sqrt{n(n+\nu)}\,\delta_{n,m+1} - \sqrt{(n+1)(n+\nu+1)}\,\delta_{n,m-1}. \tag{B2}$$

For the second case, we use results from the work of Srivastava *et al*. [13]. From Eq. (13) in [13], we obtain the following

$$\begin{aligned} F^{-}_{n,m}(k,\tau) = A_n A_m \frac{\Gamma(\nu-k+1)}{\left(1+\tau\right)^{\nu-k+1}} \sum_{j=0}^{\min(n,m)} \Bigg[ & \frac{(\nu-k+1)_j}{j!} \frac{(j+\nu+1)_{n-j}}{(n-j)!} \frac{(j+\nu+1)_{m-j}}{(m-j)!} \\ & \times \left(1+\tau\right)^{-2j} {}_2F_1\left( \left.{ -n+j, \nu-k+j+1 \atop \nu+j+1 } \right| \tfrac{1}{1+\tau} \right) {}_2F_1\left( \left.{ -m+j, \nu-k+j+1 \atop \nu+j+1 } \right| \tfrac{1}{1+\tau} \right) \Bigg] \end{aligned} \tag{B3}$$

For a proper numerical computation, we evaluate $\frac{(a)_n}{n!} = \prod_{m=1}^{n}\left(1+\frac{a-1}{m}\right)$ with $\frac{(a)_0}{0!} = 1$ and note that $(1)_n = n!$ and $(2)_n = (n+1)!$. A special case of (B3) for $\tau = 0$ and $k \neq 0$ reads as follows

$$F^{-}_{n,m}(k,0) = A_n A_m \Gamma(\nu-k+1) \sum_{j=0}^{\min(n,m)} \frac{(\nu-k+1)_j}{j!} \frac{(k)_{n-j}}{(n-j)!} \frac{(k)_{m-j}}{(m-j)!}, \tag{B4}$$

where we have used Gauss theorem ${}_2F_1\left( \left.{ a,b \atop c } \right| 1 \right) = \frac{\Gamma(c)\Gamma(c-a-b)}{\Gamma(c-a)\Gamma(c-b)}$ for $c > a+b$ and with the special case $F^{-}_{n,m}(0,0) = F^{+}_{n,m}(0,0) = \delta_{n,m}$. The result (B4) could also be obtained independently by using Eq. (17) of [13] in the evaluation of the integral (19). Moreover, using the relation ${}_2F_1\left( \left.{ a,b \atop b } \right| z \right) = \left(1-z\right)^{-a}$, we can write the following special case of (B3) with $\tau \neq 0$

$$F^{-}_{n,m}(0,\tau) = A_n A_m \frac{\tau^{n+m}\Gamma(\nu+1)}{\left(1+\tau\right)^{n+m+\nu+1}} \sum_{j=0}^{\min(n,m)} \left[ \frac{(\nu+1)_j}{j!} \frac{(j+\nu+1)_{n-j}}{(n-j)!} \frac{(j+\nu+1)_{m-j}}{(m-j)!} \frac{1}{\tau^{2j}} \right]. \tag{B5}$$

## Tables Captions

**Table 1**: A list of exactly solvable problems in the conventional formulation of quantum mechanics with a mix of continuous and discrete energy spectra such that the latter is finite. Note that the full hyperbolic Eckart potential reads $\frac{2}{\lambda^2}V(x)=\frac{A(A-1)}{\sinh^2(\lambda x)}-\frac{2B}{\tanh(\lambda x)}+2B$ and its energy spectrum is $E_k=-\frac{1}{2}\lambda^2\left[(k+A)^2+B^2(k+A)^{-2}\right]$.

**Table 2**: The parameters of orthonormal set of Laguerre bases given by Eq. (16) that are relevant to the examples in section 3. The normalization constant is $A_n=\sqrt{\Gamma(n+1)/\Gamma(n+\nu+1)}$ and $\gamma>0$.

**Table 2**: The parameters of orthonormal set of Jacobi bases given by Eq. (29) that are relevant to future studies. The normalization constant is $A_n=\sqrt{\frac{2n+\mu+\nu+1}{2^{\mu+\nu+1}}\frac{\Gamma(n+1)\Gamma(n+\mu+\nu+1)}{\Gamma(n+\mu+1)\Gamma(n+\nu+1)}}$.

## Figures Captions

**Fig. 1**: The potential function (including the orbital term) produced for the 3D system described in subsection 3.1 is shown as solid red curve. The physical parameters are taken as $\lambda=1$, $\ell=3$ and $\mu=-3.2$. The polynomial parameters are $a=b=1-\mu$. A perfect match with the function $\frac{\ell(\ell+1)}{2x^2}+\lambda V_1 x^{-1}+\lambda^3 V_2 x+V_0$ is shown by the black dots for a given fitting parameters $\{V_i\}$. Energy levels of the four bound states are shown by the horizontal black lines.

**Fig. 2**: The potential function (including the orbital term) produced for the 3D system described in subsection 3.2 is shown as solid red curve. The physical parameters are taken as $\lambda=1$, $\ell=2$ and $\mu=-4.2$. The polynomial parameters are $a=b=1-\mu$. A perfect match with the function $\frac{\ell(\ell+1)}{2x^2}+\lambda^2 V_1 x^2+V_0$ is shown by the black dots for a given fitting parameters $\{V_i\}$. Energy levels of the five bound states are shown by the horizontal black lines.

**Fig. 3**: The potential function produced for the 1D system described in subsection 3.3 is shown as solid red curve. The physical parameters are taken as $\lambda=5$, $\gamma=2$ and $\mu=-4.2$. The polynomial parameters are $a=b=1-\mu$ and the basis parameter is taken $\nu=1-2\mu$. A perfect match with the function $V_0+V_1\ln(1+\lambda x)$ is shown by the black dots for a given fitting parameters $\{V_i\}$. Energy levels of the five bound states are shown by the horizontal black lines.

**Fig. 4**: The potential function produced for the 1D system described in subsection 3.4 is shown as solid red curve. The physical parameters are taken as $\lambda=1$ and $A=-5.2$. The polynomial parameters are $a=b=1-\mu$ and the basis parameter is taken $\nu=1-2\mu$. A perfect match with the exact potential function $V(x)=\frac{1}{2}\lambda^2\left(\frac{1}{4}e^{2\lambda x}+Ae^{\lambda x}\right)$ is clear. Energy levels of the five bound states are shown by the horizontal black lines.

**Table 1**

| Potential Name | $2V(x)/\lambda^2$ | $x$ | $E_k$ |
|---|---|---|---|
| Special Eckart | $\dfrac{A(A-1)}{\sinh^2(\lambda x)}$ | $x \geq 0$ | $-\dfrac{\lambda^2}{2}(k+A)^2$ |
| 1D Morse | $\dfrac{1}{4}e^{2\lambda x} + Ae^{\lambda x}$ | $-\infty < x < +\infty$ | $-\dfrac{\lambda^2}{2}\left(k+\frac{1}{2}+A\right)^2$ |
| Pöschl-Teller | $\dfrac{A\left(A-\frac{1}{2}\right)}{\sinh^2(\lambda x/2)} - \dfrac{B\left(B-\frac{1}{2}\right)}{\cosh^2(\lambda x/2)}$ | $x \geq 0$ | $-\dfrac{\lambda^2}{2}(k+A+B)^2$ |
| 1D Scarf | $\dfrac{B^2 - A(A-1) - B(2A-1)\sinh(\lambda x)}{\cosh^2(\lambda x)}$ | $-\infty < x < +\infty$ | $-\dfrac{\lambda^2}{2}(k+A)^2$ |
| Rosen-Morse | $\dfrac{B^2 + A(A-1) + B(2A-1)\cosh(\lambda x)}{\sinh^2(\lambda x)}$ | $x \geq 0$ | $-\dfrac{\lambda^2}{2}(k+A)^2$ |

**Table 2**

| $y(x)$ | $x$ | $2\alpha$ | $2\beta$ | Normalization |
|---|---|---|---|---|
| $\lambda x$ | $x \geq 0$ | $\nu$ | 1 | $A_n$ |
| $(\lambda x)^2$ | $x \geq 0$ | $\nu + \frac{1}{2}$ | 1 | $\sqrt{2} A_n$ |
| $\gamma \ln(1 + \lambda x)$ | $x \geq 0$ | $\nu$ | $1 + \gamma^{-1}$ | $\sqrt{\gamma} A_n$ |
| $e^{\lambda x}$ | $-\infty < x < +\infty$ | $\nu + 1$ | 1 | $A_n$ |

**Table 3**

| $y(x)$ | $x$ | $2\alpha$ | $2\beta$ | Normalization |
|---|---|---|---|---|
| $1 - 2e^{-\lambda x}$ | $x \geq 0$ | $\mu + 1$ | $\nu$ | $A_n$ |
| $\tanh(\lambda x)$ | $-\infty < x < +\infty$ | $\mu + 1$ | $\nu + 1$ | $A_n$ |
| $\frac{(\lambda x)^2 - 1}{(\lambda x)^2 + 1}$ | $x \geq 0$ | $\mu + \frac{3}{2}$ | $\nu + \frac{1}{2}$ | $A_n$ |
| $2\tanh^2(\lambda x) - 1$ | $x \geq 0$ | $\mu + 1$ | $\nu + \frac{1}{2}$ | $\sqrt{\sqrt{2}} A_n$ |

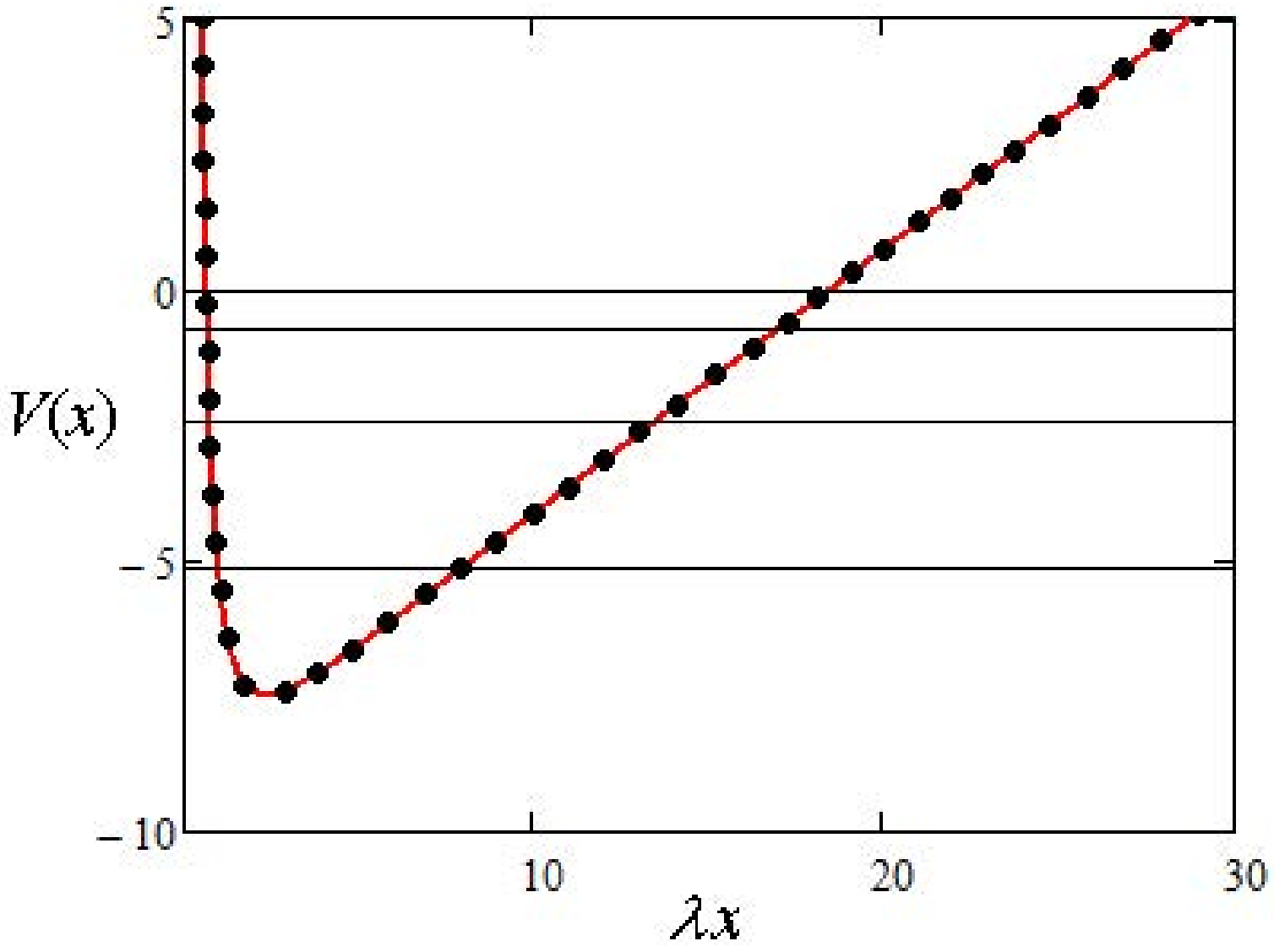

**Fig. 1**

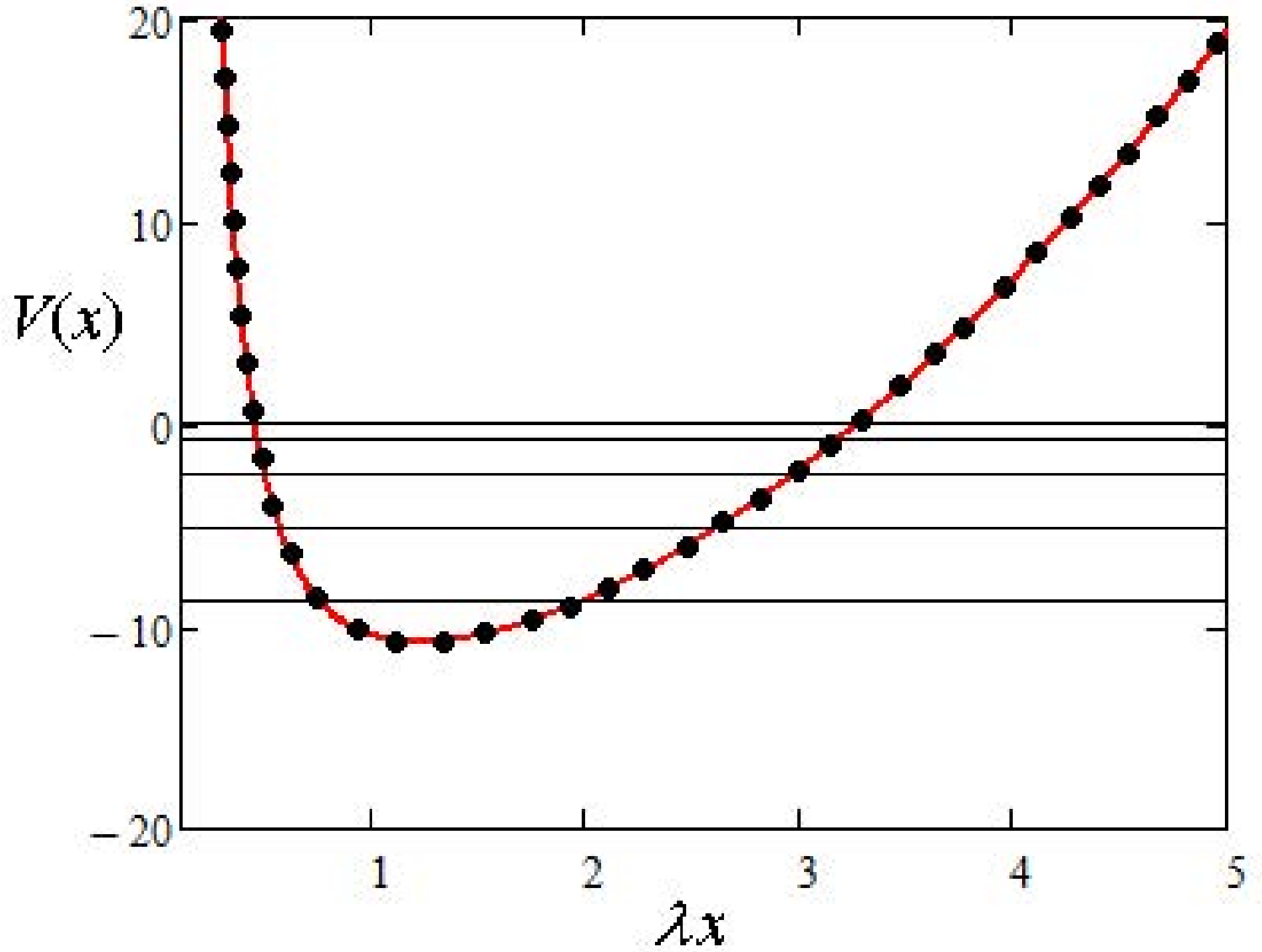

**Fig. 2**

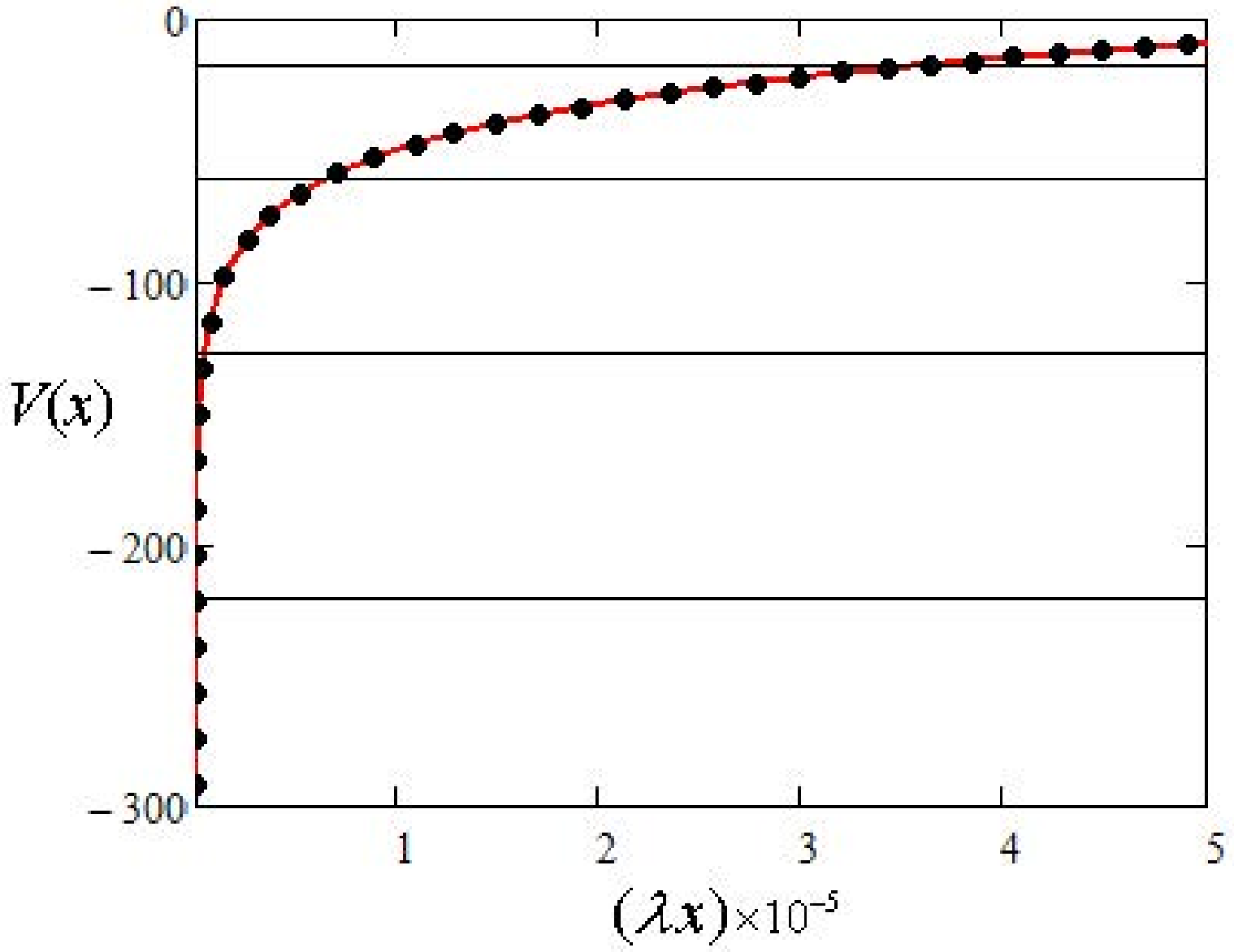

**Fig. 3**

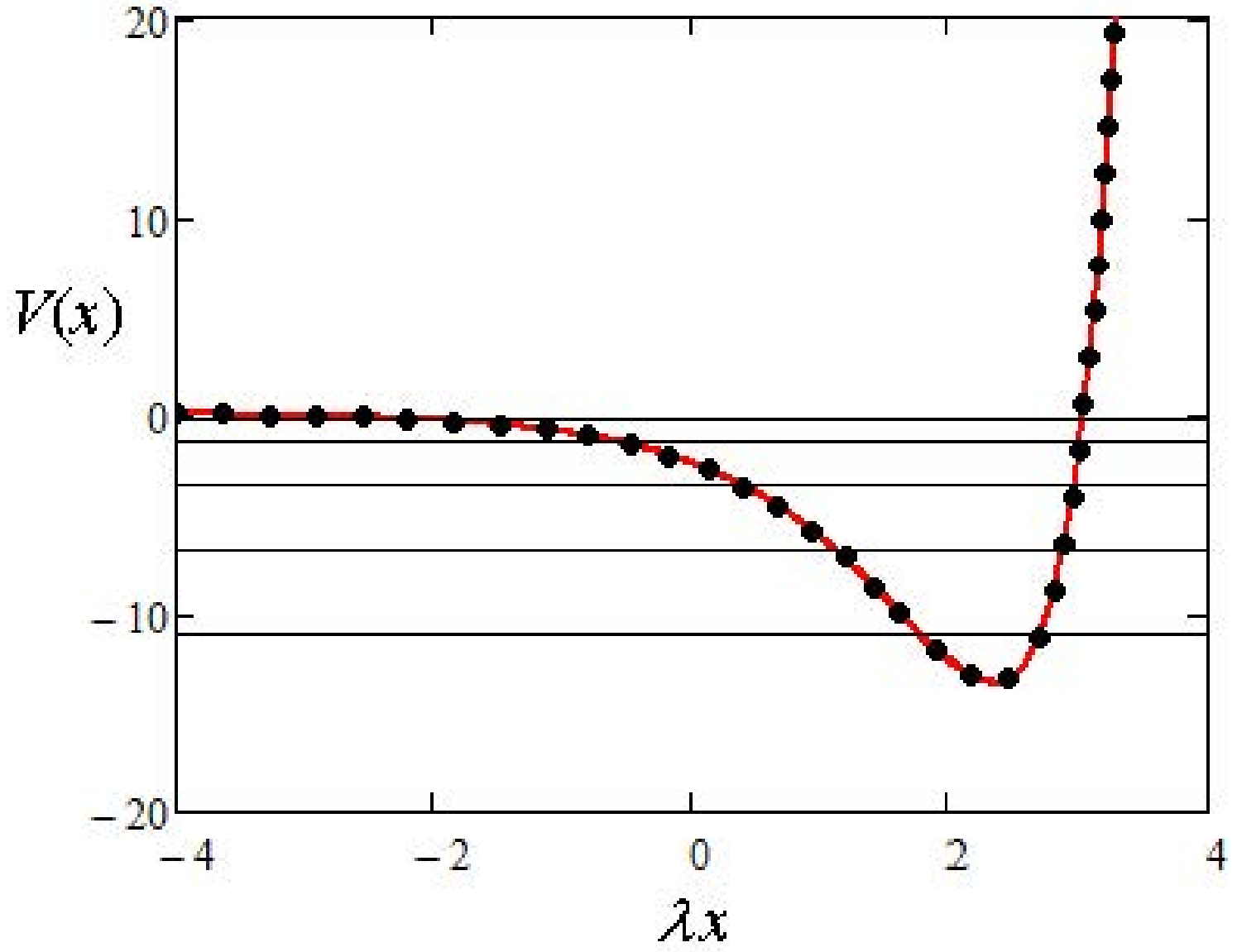

**Fig. 4**